# First and second order magnetic anisotropy and damping of europium iron garnet under high strain

*Víctor H. Ortiz[1], Bassim Arkook[1], Junxue Li[1], Mohammed Aldosary[1], Mason Biggerstaff[1], Wei Yuan[1], Chad Warren[2], Yasuhiro Kodera[2], Javier E. Garay[2], Igor Barsukov[1]\*, and Jing Shi[1]\**

[1]Department of Physics and Astronomy, University of California, Riverside, CA 92521, USA

[2] Department of Mechanical and Aerospace Engineering, University of California, San Diego, CA 92093, USA

Understanding and tailoring static and dynamic properties of magnetic insulator thin films is important for spintronic device applications. Here, we grow atomically flat epitaxial europium iron garnet (EuIG) thin films by pulsed laser deposition on (111)-oriented garnet substrates with a range of lattice parameters. By controlling the lattice mismatch between EuIG and the substrates, we tune the strain in EuIG films from compressive to tensile regime, which is characterized by X-ray diffraction. Using ferromagnetic resonance, we find that in addition to the first-order perpendicular magnetic anisotropy which depends linearly on the strain, there is a significant second-order one that has a quadratic strain dependence. Inhomogeneous linewidth of the ferromagnetic resonance increases notably with increasing strain, while the Gilbert damping parameter remains nearly constant ($\approx 2\times10^{-2}$). These results provide valuable insight into the spin dynamics in ferrimagnetic insulators and useful guidance for material synthesis and engineering of next-generation spintronics applications.

\*: Corresponding authors: Igor Barsukov (igorb@ucr.edu) and Jing Shi (jing.shi@ucr.edu)

Ferrimagnetic insulators (FMIs) have played an important role in uncovering a series of novel spintronic effects such as spin Seebeck effect (SSE) and spin Hall magnetoresistance (SMR). In addition, FMI thin films have proved to be an excellent source of proximity-induced ferromagnetism in adjacent layers (e.g., heavy metals [1], graphene [1] and topological insulators [2]) and of pure spin currents [3–6]. FMIs have also been shown to be a superb medium for magnon spin currents with a long decay length [7,8]. Among FMIs, rare earth iron garnets (REIGs) in particular have a plethora of desirable properties for practical applications: high Curie temperature ($T_c$ > 550 K), strong chemical stability, and relatively large band gaps (~ 2.8 eV).

Compared to other magnetic materials, REIGs are distinct owing to their magnetoelastic effect with the magnetostriction coefficient ranging from $-8.5 \times 10^6$ to $+21 \times 10^6$ at room temperature [9] and up to two orders of magnitude increases at low temperatures [10]. This unique feature allows for tailoring magnetic anisotropy in REIG thin films via growth, for example, by means of controlling lattice mismatch with substrates, film thickness, oxygen pressure, and chemical substitution. In thin films, the magnetization usually prefers to be in the film plane due to magnetic shape anisotropy; however, the competing perpendicular magnetic anisotropy (PMA) can be introduced by utilizing magneto-crystalline anisotropy or interfacial strain, both of which have been demonstrated through epitaxial growth [11–14]. In the study of $Tb_3Fe_5O_{12}$ (TbIG) and $Eu_3Fe_5O_{12}$ (EuIG) thin films, the PMA field $H_{2\perp}$ was found to be as high as 7 T under interfacial strain [11], much stronger than the demagnetizing field. While using strain is proven to be an effective way of manipulating magnetic anisotropy, it often comes at a cost of increasing magnetic inhomogeneity and damping of thin films [15,16].

In this work, we investigate the effect of strain on magnetic properties of (111)-oriented EuIG thin films for the following reasons: (1) The spin dynamics in EuIG bulk crystals is particularly interesting but has not been studied thoroughly in the thin film form. Compared to other REIGs, the $Eu^{3+}$ ions occupying the dodecahedral sites (c-site) should have the $J = 0$ ground state according to the Hund's rules, which do not contribute to the total magnetic moment; therefore, EuIG thin films can potentially have a ferromagnetic resonance (FMR) linewidth as narrow as that of $Y_3Fe_5O_{12}$ (YIG) [17,18] or $Lu_3Fe_5O_{12}$ (LuIG) [19]. In EuIG crystals, a very narrow linewidth (< 1 Oe) [20] was indeed observed at low temperatures, but it showed a nearly two orders of magnitude increase at high temperatures, which raises fundamental questions regarding the damping mechanism responsible for this precipitous change. (2) Although it has been shown that the uniaxial anisotropy can be controlled by moderate strain for different substrate orientations and even in polycrystalline form [21], the emergence of the higher-order anisotropy at larger strain, despite its technological significance, has remained elusive.

We grow EuIG films by pulsed laser deposition (PLD) from a target densified by powders synthesized using the method described previously [22]. The films are deposited on (111)-oriented $Gd_3Sc_2Ga_3O_{12}$ (GSGG), $Nd_3Ga_5O_{12}$ (NGG), $Gd_{2.6}Ca_{0.4}Ga_{4.1}Mg_{0.25}Zr_{0.65}O_{12}$ (SGGG), $Y_3Sc_2Ga_3O_{12}$ (YSGG), $Gd_3Ga_5O_{12}$ (GGG), $Tb_3Ga_5O_{12}$ (TGG) and $Y_3Al_5O_{12}$ (YAG) single crystal substrates, with the lattice mismatch $\eta = \frac{a_{substrate} - a_{EuIG}}{a_{EuIG}}$ (where $a$ represents the lattice parameter of the referred material) ranging from +0.45% (GSGG) to -3.95% (YAG) in the decreasing order (see Table I). After the standard solvent cleaning process, the substrates are annealed at 220 °C inside the PLD chamber with the base pressure lower than $10^{-6}$ Torr for 5 hours prior to deposition. Then the temperature is increased to ~ 600 °C in the atmosphere of 1.5 mTorr oxygen mixed with 12% (wt.) ozone for 30 minutes. A 248 nm KrF excimer pulsed laser is used to ablate the target with a power of 156 mJ and a repetition rate of 1 Hz. We crystalize the films by *ex situ* annealing at 800 °C for 200 s in a steady flow of oxygen using rapid thermal annealing (RTA).

Reflection high energy electron diffraction (RHEED) is used to evaluate the crystalline structural properties of the EuIG films grown on various substrates (Fig. 1a). Immediately after the deposition, RHEED displays the absence of any crystalline order. After *ex situ* rapid thermal annealing, all EuIG films turn into single crystals. We carry out atomic force microscopy (AFM) on all samples and find that they show atomic flatness and good uniformity with root-mean-square (RMS) roughness < 2 Å (Fig. 1b). In addition, we perform X-ray diffraction (XRD) on all samples using a Rigaku SmartLab with Cu Kα radiation with a Ni filter and Ge(220) mirror as monochromators, at room temperature in 0.002° steps over the $2\theta$ range from 10° to 90° [23]. In a representative XRD spectrum (Fig. 1c), two (444) Bragg peaks are present, one from the 50 nm thick EuIG film and the other from the YSGG substrate, which confirms the epitaxial growth and single crystal structure of the film without evidence of any secondary phases. Other REIG films grown under similar conditions, i.e., by PLD in oxygen mixed with ozone at ~600 °C during followed by RTA, have shown no observable interdiffusion across the interface from high resolution transmission electron microscopy and energy dispersive X-ray spectroscopy (Fig. S1, [24]). The EuIG Bragg peak ($a_0 = 12.497$ Å) is shifted with respect to the expected peak position of unstrained bulk crystal, indicating a change in the EuIG lattice parameter perpendicular to the surface ($a_\perp$). For the example shown in Fig. 1c, the EuIG (444) peak shifts to left with respect to its bulk value, indicating an out-of-plane tensile strain and therefore an in-plane compressive strain in the EuIG lattice.

A common approach for inferring the in-plane strain $\varepsilon_\parallel$ of thin films from the standard $\theta$–$2\theta$ XRD measurements involves the following equation [23],

$$\varepsilon_\parallel = -\frac{c_{11} + 2\,c_{12} + 4\,c_{44}}{2c_{11} + 4\,c_{12} - 4\,c_{44}}\,\varepsilon_\perp, \qquad \text{with } \varepsilon_\perp = \frac{a_\perp - a_o}{a_o}, \tag{1}$$

where $a_0$ is the lattice parameter of the bulk material, and $a_\perp$ can be calculated using $a_\perp = d_{hkl}\sqrt{h^2 + k^2 + l^2}$ from the interplanar distance $d_{hkl}$ obtained from the XRD data (Fig. S2, [25]), and $c_{ij}$ are the elastic stiffness constants of the crystal which in most cases can be found in the literature [9]. However, due to the wide range of strain values studied in this work and the possibility that the films may contain different amounts of crystalline defects, we perform reciprocal space mapping (RSM) measurements on a subset of our EuIG samples (Fig. S3, [26]) and compared the measured in-plane lattice parameters with the calculated ones using Eq. 1. We observe that the average in-plane strains measured by RSM has a systematic difference of 40% from the calculated values based on the elastic properties (Fig. S4, [26]). Given this nearly constant factor for all measured films, we find that the elastic stiffness constants of our EuIG films may deviate from the literature reported bulk values, possibly due to stochiometric deviations or slight unit cell distortion in thin films. Here we adopt the reported lattice parameter value ($a_0$ = 12.497 Å) as the reference due to the difficulty of growing sufficiently thick, unstrained EuIG films using PLD.

In the thickness-tuned magnetic anisotropy study [11], the anisotropy field in REIG films is found to be proportional to $\eta/(t+t_o)$, which was attributed to the relaxation of strain as the film thickness $t$ increases. Here in EuIG samples with small lattice mismatch $\eta$ (e.g., NGG/EuIG), the strain is mostly preserved in 50 nm thick films (pseudomorphic regime), whereas for larger $\eta$ (e.g., YAG/EuIG), the lattice parameter of EuIG films shows nearly complete structural relaxation to the bulk value. For this reason, in the samples with larger $\eta$ (YAG = -3.95 %, GSGG = 0.45%), we grow thinner EuIG films (20 nm) in order to retain a larger in-plane strain (compressive for YAG, tensile for GSGG). For EuIG films grown on TGG and GGG substrates, the paramagnetic background of the substrates is too large to obtain a reliable magnetic moment measurement of the EuIG films; therefore, the results of thinner films on these two substrates are not included in this study.

Room-temperature magnetic hysteresis curves for YSGG/EuIG sample are shown in Fig. 1d with the magnetic field applied parallel and perpendicular to the film [26]. The saturation field for the out-of-plane loop (~1100 Oe) is clearly larger than that for the in-plane loop, indicating that the magnetization prefers to lie in the film plane. Moreover, since the demagnetizing field $4\pi M_s$ ($\approx$ 920 Oe) is less than the saturation field in the out-of-plane loop (Fig. S5, [27]), it suggests the presence of additional easy-plane anisotropy resulting from the magnetoelastic effect due to interfacial strain. As shown in this example, we can qualitatively track the evolution of the magnetic anisotropy in samples with different strains. However, this approach cannot provide a quantitative description when high-order anisotropy contributions are involved.

To quantitatively determine magnetic anisotropy in all EuIG films, we perform polar angle ($\theta_H$)-dependent FMR measurements using an X-band microwave cavity with frequency $f$ = 9.32 GHz and field modulation. The samples are rotated from $\theta_H$ = 0° to $\theta_H$ = 180° in 10° steps, where $\theta_H$ = 90° corresponds to the field parallel to the sample plane (Fig. 2a). The spectra at $\theta_H$ = 0° for all samples are displayed in Fig. 2b and show a single resonance peak which can be well fitted by a Lorentzian derivative. Despite different strains in all samples, the resonance field $H_{res}$ is lower for the in-plane direction ($\theta_H$ = 90°) than for the out-of-plane direction ($\theta_H$ = 0°). A quick inspection reveals that the out-of-plane $H_{res}$ shifts to larger values as $\eta$ increases in the positive direction (e.g., from YAG/EuIG to GSGG/EuIG), corresponding to stronger easy-plane anisotropy. Furthermore, the $H_{res}$ values at $\vartheta_H$ = 0° show a large spread among the samples. Fig. 2c shows a comparison of FMR spectra at different polar angles between two representative samples: NGG/EuIG (small $\eta$) and YAG/EuIG (large $\eta$).

Figs. 3a-c show $H_{res}$ vs. $\vartheta_H$ for three representative EuIG films. To evaluate magnetic anisotropy, we fit the data using the Smit-Beljers formalism by considering the first-order $-K_1 \cos^2 \theta$ and the second-order $-\frac{1}{2} K_2 \cos^4 \theta$ uniaxial anisotropy energy terms [28]. From this fitting, we extract the parameters $4\pi M_{eff} = 4\pi M_s - \frac{2K_1}{M_s} = 4\pi M_s - H_{2\perp}$ and $H_{4\perp} = \frac{2K_2}{M_s}$ (see Table I), here $H_{2\perp}$ and $H_{4\perp}$ being the first- and second- order anisotropy fields, respectively, and favoring

out-of-plane (in-plane) orientation of magnetization when they are positive (negative). The spectroscopic $g$-factor is treated as a fitted parameter which is found as a nearly constant, $g = 1.40$ (Fig. S6, [28]), in accordance to the previous results obtained by Miyadai [31]. In Figs. 3d and 3e, we present $H_{2\perp}$ and $H_{4\perp}$ as functions of the measured out-of-plane strain $\varepsilon_\perp$ and in-plane strain $\varepsilon_\parallel$. Clearly, the magnitude of $4\pi M_{eff}$ is greater than the demagnetizing field for EuIG $4\pi M_s = 920\ Oe$; therefore, $H_{2\perp}$ is negative for all samples, i.e., favoring the in-plane orientation. As shown in Fig. 3d, $|H_{2\perp}|$ increases linearly with increasing in-plane strain $\eta$. This is consistent with the magnetoelastic effect in (111)-oriented EuIG films [9]. As briefly discussed earlier, due to the constant scaling factor between the calculated and measured $\varepsilon_\parallel$, we rewrite the magnetoelastic contribution to the first-order perpendicular anisotropy as $-\frac{9\Xi}{3M_s}\varepsilon_\perp$, with the parameter $\Xi$ containing the information related to the magnetoelastic constant $\lambda_{111}$ and elastic stiffness $c_{ii}$. We fit the magnetoelastic equation in Ref. [11] using the parameter $\Xi$ and obtain $\Xi = -(7.06 \pm 0.95) \times 10^4\ \frac{dyne}{cm^2}$ from the slope. On the other hand, based on the reported literature values ($\lambda_{111} = +1.8 \times 10^{-6}$, $c_{11} = 25.10 \times 10^{11}$ dyne/cm$^2$, $c_{12} = 10.70 \times 10^{11}\ \frac{dyne}{cm^2}$, $c_{44} = 7.62 \times 10^{11}\ \frac{dyne}{cm^2}$) [10], we obtain $\Xi_{lit} = -6.12 \times 10^4\ \frac{dyne}{cm^2}$. This result suggests that even though the actual elastic properties of our EuIG films may be different from the ones reported in for EuIG crystals due to the thin film unit cell distortion (Table S1, [32]), the pertaining parameter $\Xi$ appears to be relatively insensitive to variations of stoichiometry. The intercept of the straight-line fit should give the magneto-crystalline anisotropy coefficient of EuIG $K_c$. We find $K_c = (+62.76 \pm 0.18) \times 10^3$ erg/cm$^3$, which is different from the previously reported values for EuIG bulk crystals in both the magnitude and sign ($K_c = -38 \times 10^3$ erg/cm$^3$) [31]. Similar growth-modified magneto-crystalline anisotropy was observed in EuIG films grown with relatively low temperatures (requiring post-deposition annealing to crystalize) [10]. In the absence of interfacial interdiffusion, the anomalous anisotropy may be related to partial deviation from the chemical ordering of the garnet structure [31].

By comparing the first- and second-order anisotropy fields $H_{2\perp}$ and $H_{4\perp}$ vs. $\varepsilon_\parallel$ plotted in Figs. 3d and 3e, we find that the former dominates over the entire range of $\varepsilon_\parallel$ (except for YAG/EuIG). In contrast to the linear dependence for $H_{2\perp}$, $H_{4\perp}$ can be fitted well with a quadratic $\varepsilon_\parallel$ dependence, which is not surprising for materials with large magnetostriction constants (such as EuIG) under large strains. For relatively small $\varepsilon_\parallel$, the linear strain term in the magnetic anisotropy energy dictates. For large $\varepsilon_\parallel$, higher-order strain terms may not be neglected. By including the $(\varepsilon_\parallel \cos^2\vartheta)^2$ term, we obtain excellent fitting to the FMR data, indicating that the second-order expansion in $\varepsilon_\parallel$ is adequate. In contrast to $H_{2\perp}$, $H_{4\perp}$ is always positive, thus favoring out-of-plane magnetization orientation. It is worth pointing out that for YAG and TGG, the magnitude of the $H_{2\perp}$ becomes comparable with that of the $H_{4\perp}$, but the sign differs. Comparison of $H_{4\perp}$ with $4\pi M_{eff}$ reveals that a coexistence (bi-stable) magnetic state can be realized when $H_{4\perp} > 4\pi M_{eff}$ [31, 33-35]. The results are summarized in Table I.

The above magnetic anisotropy energy analysis only deals with the polar angle dependence, but in principle, it can also vary in the film plane and therefore depend on the azimuthal angle. To understand the latter, we perform azimuthal angle dependent FMR measurements on all samples. We indeed observe a six-fold in-plane anisotropy in $H_{res}$ due to the crystalline symmetry of EuIG (111). However, the amplitude of the six-fold $H_{res}$ variation is less than 15 Oe, about two orders of magnitude smaller than the average value of $H_{res}$ for most samples, thus we omit the in-plane anisotropy in our analysis.

Besides the $H_{res}$ information, the FMR spectra in Fig. 2c reveals significant variations in FMR linewidth, which contains information of magnetic inhomogeneity and Gilbert damping. To investigate these properties systematically, we perform broad-band (up to 15 GHz) FMR measurements with magnetic field applied in the film plane, using a coplanar waveguide setup. From the frequency dependence of $H_{res}$, we obtain $4\pi M_{eff}$ and $g$ independently via fitting the data with the Kittel equation. These values agree very well with those previously found from the polar angle dependence. We plot the half width at half maximum, $\Delta H$, as a function of frequency $f$ in Fig. 4a. While $\Delta H$ varies significantly across the samples, the data for each sample fall approximately on a straight line and the slope of $\Delta H$ vs. $f$ appears to be visibly close to each other. For a quantitative evaluation of $\Delta H$, we consider the following contributions: the Gilbert damping $\Delta H_{Gilbert}$, two-magnon scattering $\Delta H_{TMS}$, and the inhomogeneous linewidth $\Delta H_0$ [36],

$$\Delta H = \Delta H_{Gilbert} + \Delta H_{TMS} + \Delta H_0 . \tag{3}$$

The Gilbert term, $\Delta H_{Gilbert} = \frac{2\pi\alpha f}{|\gamma|}$, depends linearly on $f$, where $\alpha$ is the Gilbert damping parameter; the two-magnon term is described through $\Delta H_{TMS} = \Gamma_0 \arcsin \sqrt{\frac{\sqrt{f^2+\left(\frac{f_o}{2}\right)^2}-\frac{f_o}{2}}{\sqrt{f^2+\left(\frac{f_o}{2}\right)^2}+\frac{f_o}{2}}}$ [37], where $\Gamma_0$ denotes the magnitude of the two-magnon scattering, $f_0 = 2\gamma M_{\text{eff}}$; and $\Delta H_0$, the inhomogeneous linewidth which is frequency independent.

By fitting Eq. (3) to the linewidth data, we obtain quantitative information on magnetic damping through the Gilbert parameter and two-magnon scattering magnitude as well as the magnetic inhomogeneity [39–40]. In Fig. 4a, the overall linear behavior for all samples is an indication of a relatively small two-magnon scattering contribution $\Delta H_{TMS}$ which therefore may be disregarded in the fitting process. Figs. 4b and 4c show both $\Delta H_0$ and $\alpha$ vs. $\varepsilon_\parallel$. It is clear that four of the samples with the smallest $\Delta H_0$ (~ 10 Oe) are those with relatively low in-plane strain ($|\varepsilon_\parallel| < 0.30\%$). In the meantime, the XRD spectra of these samples show fringes characteristic of well conformed crystal planes (Fig. S2), and moreover, the RSM plots (Fig. S3) reveal a uniform strain distribution in the films [41]. On the compressive strain side, $\Delta H_0$ increases steeply to 400 Oe at $\varepsilon_\parallel \sim -0.40 \%$, and their XRD spectra show no fringes and the RSM graphs indicate non-uniform strain relaxation in the samples (Figs. S2 and S3). In sharp contrast to the $\Delta H_0$ trend, the Gilbert damping $\alpha$ remains about $2\times10^{-2}$ over the entire range of $\varepsilon_\parallel$, suggesting that the intrinsic magnetic damping of EuIG films is nearly unaffected by the inhomogeneity. In fact, the magnitude of $\alpha$ is significantly larger than that of YIG [17,18] or LuIG films [19], which is somewhat unexpected for $Eu^{3+}$ in EuIG with $J = 0$. A possible reason for this enhanced damping is that other valence states of Eu such as $Eu^{2+}$ ($J = 7/2$) may be present, which leads to non-zero magnetic moments of Eu ions in the EuIG lattice and thus results in a larger damping constant, common to other REIG with non-zero 4f-moments [42]. The X-ray photoelectron spectroscopy data taken on YSGG(111)/EuIG(50 nm) (Fig. S7, [43]) indicates such a possibility. While the FMR linewidth presents large variations across the sample set, we have identified that the non-uniform strain relaxation process caused by large lattice mismatch with the substrate is a main source of the inhomogeneity linewidth $\Delta H_0$, but it does not affect the Gilbert damping $\alpha$. The results raise interesting questions on the mechanisms of intrinsic damping and the origin of magnetic inhomogeneity in EuIG thin films, both of which warrant further investigations.

In summary, we find that uniaxial magnetic anisotropy in PLD-grown EuIG(111) thin films can be tuned over a wide range via magnetostriction and lattice-mismatch induced strain. The first-order anisotropy field depends linearly on the strain and the second order anisotropy field has a quadratic dependence. While non-uniform strain relaxation significantly increases the magnetic inhomogeneity, the Gilbert damping remains nearly constant over a wide range of in-plane strain. The results demonstrate broad tunability of magnetic properties in REIG films and provide guidance for implementation of EuIG for spintronic applications. Further studies to elucidate the role of $Eu^{2+}$ sites in magnetic damping are called upon.

We thank Dong Yan and Daniel Borchardt for their technical assistance. This work was supported as part of the SHINES, an Energy Frontier Research Center funded by the US Department of Energy, Office of Science, Basic Energy Sciences under Award No. SC0012670. J.S. acknowledges support by DOE BES Award No. DE-FG02-07ER46351 and I.B. acknowledges support by the National Science Foundation under grant number NSF-ECCS-1810541.

# Figures

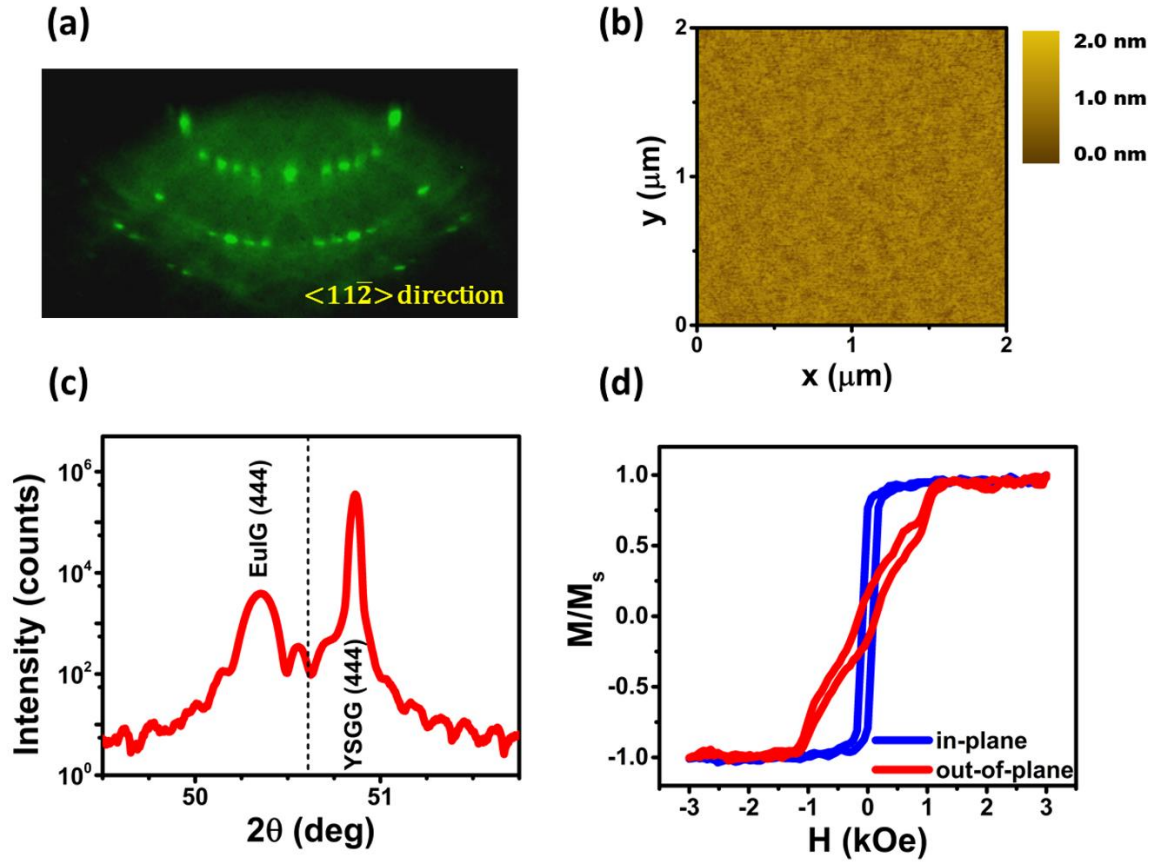

**Figure 1**: Structural and magnetic property characterization of EuIG 50 nm film grown on TGG(111) substrate. (a) Reflection high energy electron diffraction (RHEED) pattern along the $\langle 11\bar{2} \rangle$ direction, displaying single crystal structure after rapid thermal annealing process. (b) 5 mm × 5 mm atomic force microscope (AFM) surface morphology scan, demonstrating a root-mean-square (RMS) roughness of 1.8 Å. (c) Intensity semi-log plot of $\theta$-$2\theta$ XRD scan. The dashed line corresponds to the XRD peak for bulk EuIG. (d) Magnetization hysteresis loops for field out-of-plane and in-plane directions.

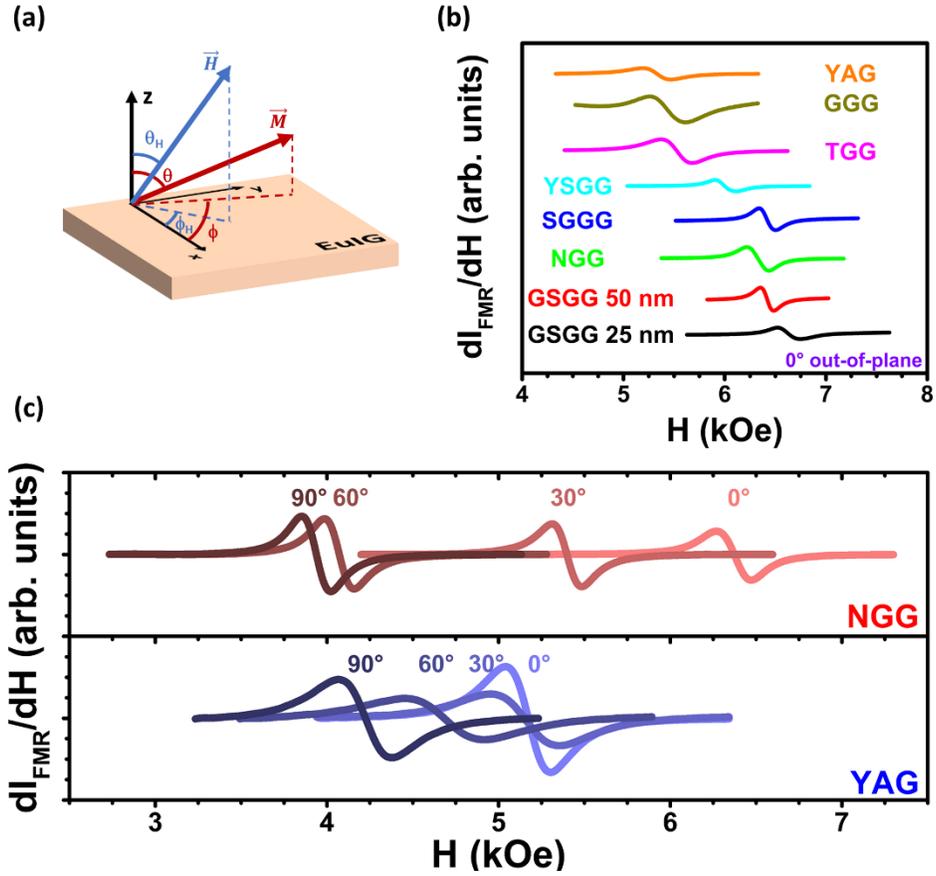

**Figure 2** Polar angle dependent ferromagnetic resonance (FMR). (a) Coordinate system used for the FMR measurement. (b) Room temperature FMR derivative absorption spectra for $\theta_H = 0°$ (out-of-plane configuration) for EuIG on different (111) substrates. (c) FMR derivative absorption spectra for 50 nm EuIG grown on NGG(111) ($\varepsilon_\parallel \approx 0$) and 20 nm EuIG on YAG(111) ($\varepsilon_\parallel < 0$) with polar angle $\vartheta_H$ ranging from 0° (out-of-plane) to 90° (in-plane) at 300 K, where $\varepsilon_\parallel$ is in-plane strain between the EuIG film and substrate.

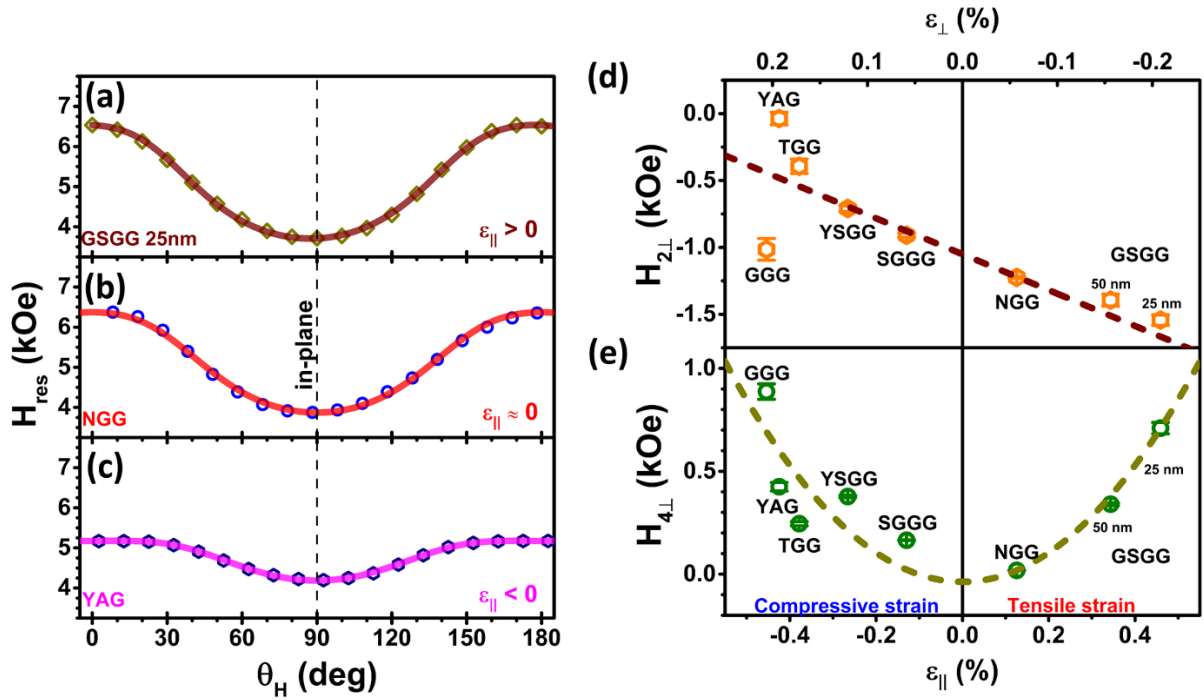

**Figure 3** Polar angle dependent ferromagnetic resonance field $H_{res}$ for (a) tensile in-plane strain ($\varepsilon_\parallel > 0$), (b) in-plane strain close to zero ($\varepsilon_\parallel \approx 0$), and (c) compressive in-plane strain ($\varepsilon_\parallel < 0$). Solid curves represent the best fitting results. In-plane strain dependence of the anisotropy fields $H_{2\perp}$ (d) and $H_{4\perp}$ (e).

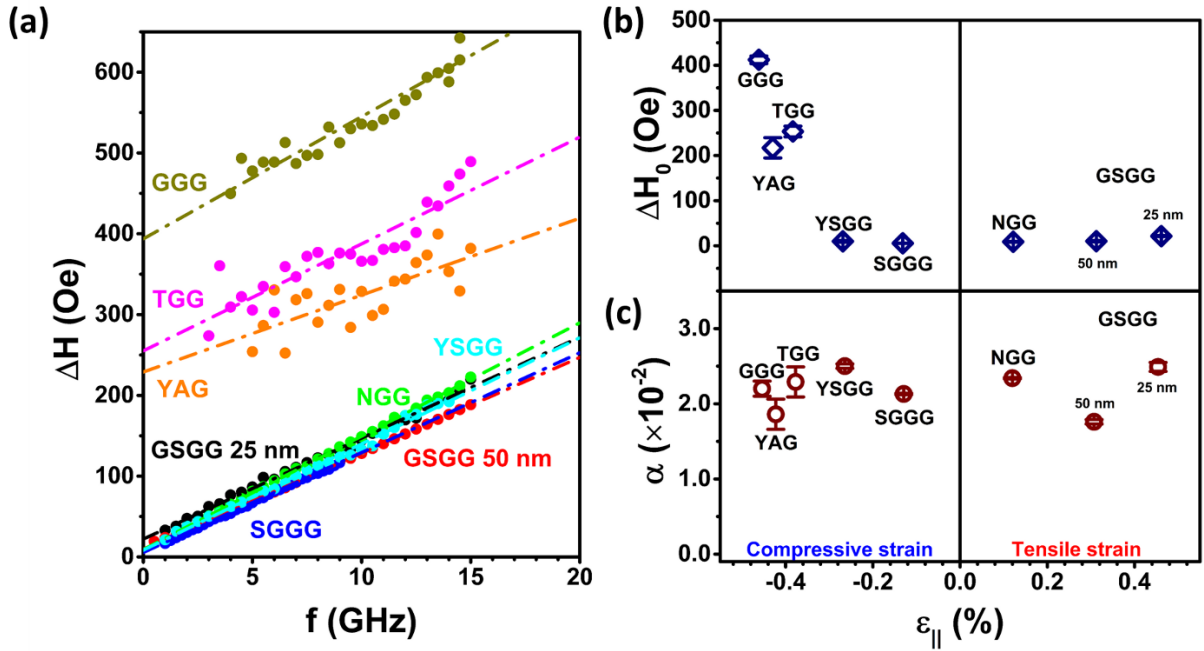

**Figure 4** FMR linewidth and magnetic damping of EuIG films as a function of in-plane strain. (a) Half width at half maximum $\Delta H$ vs. frequency $f$ for EuIG films grown on different substrates, with the corresponding fitting according to Eq. (3). In-plane strain dependence of inhomogeneous linewidth $\Delta H_0$ (b) and Gilbert parameter α (c).

| Substrate | $a_{\text{substrate}}$ (Å) | $\eta$ (%) | $t$ (nm) | $\varepsilon_\parallel$ (%) | $\varepsilon_\perp$ (%) | $g$ | $H_{2\perp}$ (Oe) | $H_{4\perp}$ (Oe) | $\alpha$ (×10$^{-2}$) | $\Delta H_o$ (Oe) | $\Gamma_o$ (Oe) |
|---|---|---|---|---|---|---|---|---|---|---|---|
| GSGG | 12.554 | 0.45 | 50 | 0.34 | -0.16 | 1.40 | -1394.2 ± 44.9 | 339.79 ± 6.59 | 2.46 ± 0.03 | 21.4 ± 1.3 | 2.61 |
|  |  |  | 25 | 0.46 | -0.21 | 1.41 | -1543.6 ± 39.7 | 709.47 ± 27.5 | 1.58 ± 0.06 | 10.2 ± 1.7 | 6.05 |
| NGG | 12.508 | 0.06 | 50 | 0.12 | -0.06 | 1.38 | -1224.4 ± 5.7 | 18.34 ± 0.05 | 2.41 ±0.01 | 8.9 ± 0.7 | 0.20 |
| SGGG | 12.480 | -0.14 | 50 | -0.13 | 0.06 | 1.40 | -909.6 ± 15.2 | 164.8 ± 1.36 | 2.13 ± 0.01 | 5.6 ± 0.4 | 0.50 |
| YSGG | 12.426 | -0.57 | 50 | -0.27 | 0.12 | 1.37 | -709.4 ± 22.0 | 377.3 ± 5.09 | 2.47 ± 0.03 | 9.9 ± 1.8 | 2.47 |
| GGG | 12.383 | -0.92 | 50 | -0.45 | 0.21 | 1.38 | -1015.0 ± 81.3 | 887.2 ± 37.27 | 2.20 ± 0.14 | 412.2 ± 8.4 | 3.35 |
| TGG | 12.355 | -1.14 | 50 | -0.38 | 0.18 | 1.38 | -393.4 ± 53.6 | 245.0 ± 10.00 | 2.29 ± 0.20 | 253.4 ± 11.8 | 0.20 |
| YAG | 12.004 | -3.95 | 20 | -0.42 | 0.20 | 1.37 | -36.8 ± 47.1 | 424.8 ± 20.91 | 1.86 ± 0.20 | 217.0 ± 22.6 | 0.20 |

**Table 1** Structural and magnetic parameters for the EuIG thin films grown on different substrates.